\documentclass{aa}  
\usepackage{graphicx,natbib}
\usepackage{txfonts}
\newcommand{\Msun}{\ensuremath{\mathit{M}_{\odot}}}

\begin{document} 

   \authorrunning{Moriya}
   \titlerunning{Mass loss of SN progenitors by core neutrino emission}

   \title{
Mass loss of massive stars near the Eddington luminosity \\
by core neutrino emission shortly before their explosion
 }

   \subtitle{}

   \author{Takashi J. Moriya
          \inst{1,2}
          }

   \institute{
Argelander Institute for Astronomy, University of Bonn,
Auf dem H\"ugel 71, D-53121 Bonn, Germany
   \\
              \email{moriyatk@astro.uni-bonn.de}\and
Research Center for the Early Universe, Graduate School of Science,
University of Tokyo, Hongo 7-3-1, Bunkyo, Tokyo, Japan
             }

   \date{Received November 6, 2013; accepted March 11, 2014}

\abstract{
We present a novel mechanism to enhance the mass-loss rates of massive
stars shortly before their explosion. The neutrino luminosities of
the stellar core of massive stars become larger as they get closer
to the time of the core collapse.
As emitted neutrinos escape freely from the core, the core mass is
significantly reduced when the neutrino luminosity is large.
If a star is near the Eddington luminosity
when the neutrino luminosity is large, the star can exceed
the Eddington luminosity because of the core neutrino mass loss.
We suggest that the stellar surface mass-loss rates due to the core
neutrino emission can be higher than $10^{-4}$ $M_\odot~\mathrm{yr^{-1}}$
from $\sim$ 1 year before the core collapse. The mass-loss rates can exceed
$10^{-2}$ $M_\odot~\mathrm{yr^{-1}}$ in $\sim 10$ days before the core
collapse.
This mass-loss mechanism may be able to explain the enhanced
mass loss observed in some supernova progenitors shortly before their
explosion.
Even if the star is not close enough to the Eddington luminosity
to enhance mass loss, the star can
still expand because of the reduced gravitational force.
This mechanism can be activated in Wolf-Rayet stars and it can
make hydrogen-poor, as well as hydrogen-rich, dense circumstellar media
observed in some supernovae.
}

 

   \keywords{stars: massive -- stars: mass-loss -- supernovae: general}

   \maketitle
%

\section{Introduction}\label{introduction}
Stellar mass loss is one of the most influential phenomena which
determine the fate of a massive star.
For instance, the spectral type of an exploding star, a supernova (SN),
depends on the amount of mass loss during the evolution.
There is a growing evidence that some stars experience
very high mass-loss rates shortly before the explosion.
SN 2006jc is, for the first time, found to have got bright two years
before its explosion as a Type Ibn SN \citep{pastorello2007}.
A similar Type Ibn SN iPTF13beo is found to have two light curve peaks
which indicate that the mass-loss rate of the Wolf-Rayet (WR) progenitor
is $\sim 2\times 10^{-3}$ $M_\odot~\mathrm{yr^{-1}}$ in a year before
the explosion \citep{gorbikov2013}.
Type IIn SNe are more commonly discovered SNe with dense
circumstellar media (CSM) resulting from the high mass-loss rates
of their progenitors shortly before their explosions.
They are typically higher than $\sim 10^{-3}$
$M_\odot~\mathrm{yr^{-1}}$ \citep[e.g.,][]{taddia2013,kiewe2012,moriya2013}.

Having such high mass-loss rates immediately before the
explosions of these SN progenitors is not expected from the standard
stellar evolution theory and its mechanisms are still not known well.
Some of them are likely to be related to the progenitors' very high
luminosities and their proximity to the Eddington luminosity.
The progenitor of Type IIn SN 2005gl is found to
have an absolute brightness of $M_V\simeq-10$ mag 
and it belongs to the most luminous class of stars.
The progenitor of Type IIn SN 2009ip provided us with more remarkable
information about the progenitors' activities
(e.g., \citealt{margutti2013,smith2013,mauerhan2013,prieto2013},
but see also \citealt{fraser2013,pastorello2013}).
The luminosity of the progenitor was, again, close to $M_V=-10$ mag and it is
presumed to be near the Eddington luminosity. The progenitor started to
show several outbursts from a few years before its explosion.
The final burst in August 2012 started about 50 days before
the major luminosity increase started in September 2012
which may be caused by the SN explosion.
A dense CSM could have been formed during these eruptive events.
Type IIn SNe 2010mc and 2011ht
also experienced a similar pre-SN outburst \citep{ofek2013,fraser2013b}.
A Type Ic superluminous SN 2006oz had a precursor from about 10 days before its
major luminosity increase \citep{leloudas2012}.
This precursor may also be originated from the progenitor activities
shortly before the explosion (but see also \citealt{moriya2012}
for another interpretation).

In addition to the unexpected mass-loss enhancement
shortly before the core collapse,
some stars may have larger radii than those expected from the stellar evolution
theory at the time of the explosion. 
The early emission of Type Ib SN 2008D is suggested to be difficult to
be explained by standard WR stars \citep[e.g.,][]{dessart2011}.
\citet{bersten2013} showed that the radius of the WR progenitor of SN
2008D should be $\sim 10~R_\odot$ to explain the early emission 
by the adiabatic cooling of the SN ejecta and
they concluded that $^{56}$Ni may have pumped out to
the outermost layers of the WR progenitor
(see also, e.g., \citealt{couch2011,balberg2011} for other
interpretations).
Some fast evolving Type Ic SNe are suggested to originate from
extended WR stars whose radii are $\sim 10~R_\odot$
(\citealt{kleiser2013}, but see also \citealt{tauris2013}).
The progenitors of ultra-long gamma-ray bursts are unlikely to
be WR stars with their typical radii because of their long durations
\citep[e.g.,][]{levan2013}.
Those observations may infer that some WR stars explode with much larger radii
than those predicted by the current stellar evolution theory.

One of the characteristics of massive stars shortly before the
explosion is their huge neutrino luminosities. 
Neutrinos are constantly emitted from the stellar core
throughout the evolution because of the
nuclear reactions.
However, after the onset of the carbon burning,
the thermal neutrino
emission becomes significant because of the high temperature required
for the ignitions of heavy elements.
The expected neutrino luminosity is so large that
dying stars will be detected by the modern neutrino detectors
from a few weeks before the core collapse
if they are close enough to the Earth
\citep{odrzywolek2011,odrzywolek2004}.
The high neutrino luminosity from the core
together with the energy release by nuclear burning and the composition
gradients results in the active convective motion which is suggested to
be related to the high mass-loss rates of SN progenitors
\citep[e.g.,][]{quataert2012,shiode2013}.
Here, in this paper, we suggest another possible effect caused by
the huge neutrino luminosity which may result in the extreme mass loss at
the stellar surface. The mass loss of the core due to the neutrino emission
leads to the sudden decrease of the gravitational force.
The effect of the mass loss due to large neutrino luminosities
after the core collapse of a star
has been discussed \citep[][]{lovegrove2013,nadezhin1980}.
In this paper, we suggest that the effect of the neutrino mass loss at the
stellar core can appear at the stellar surface in massive stars near the
Eddington luminosity immediately before their core collapse.
In short, the gravitational potential can be significantly reduced
because of the neutrino emission shortly before the core collapse
and the stellar luminosity can exceed the Eddington luminosity.
Thus, the weakened gravitational force
results in the enhanced mass loss.
It may also result in the expansion of massive stars
shortly before their explosion.

\section{Neutrino mass loss from stellar core}
At the very late stage of the stellar evolution, especially after 
the onset of the core carbon burning, the neutrino luminosity from the
stellar core starts to be dominated by thermal neutrinos.
The emerged neutrino freely escapes from the core.
The mass-loss rate $\dot{M}_c$ from the core with the neutrino
luminosity $L_\nu$ can be estimated from a simple relation $L_\nu=\dot{M}_cc^2$,
\begin{eqnarray}
\dot{M}_c&=&\frac{L_\nu}{c^2}, \\
&=&6.8\times 10^{-3}\left(\frac{L_\nu}{10^{11}L_\odot}\right)~M_\odot~\mathrm{yr^{-1}},
\label{neumasslossrate}
\end{eqnarray}
where $c$ is the speed of light.

To estimate the mass-loss rate from the stellar core by the neutrino emission,
we numerically follow stellar evolution by using a public stellar
evolution code \texttt{MESAstar} version 5527 \citep{paxton2011,paxton2013}.
We evolve solar metallicity ($Z=0.02$) non-rotating stars
whose zero-age main-sequence (ZAMS) mass is 25 \Msun\ and 50 $M_\odot$.
We evolve the stars until the core collapse.
The mixing-length parameter is set as $\alpha_\mathrm{MLT}=1.6$.
We use the MLT++ module for the convection in order to follow the
evolution up to the core collapse.
The Ledoux criterion is adopted for the convection with the
dimension-less semiconvection efficiency of $\alpha_\mathrm{sc}=1.0$
without overshooting.
The wind mass loss from the stellar surface is based on \citet{vink2001} for
the stars with the surface effective temperature above $10^4$ K.
The mass-loss rates are reduced by 0.8 from the original \citet{vink2001}
prescription \citep{maeder2001}.
For the lower temperatures, we adopt the prescription by \citet{dejager1988}.
Neutrino emission in the code is evaluated based on \citet{itoh1996}.
At the time of the core collapse, the 25 \Msun\ star has evolved
to a 20 \Msun\ red supergiant with a 4 \Msun\ carbon+oxygen core, while
the 50 \Msun\ star has become a 22 \Msun\ star with a 15 \Msun\
carbon+oxygen core and a helium-rich layer outside.

\begin{figure}
\centering
\includegraphics[width=\columnwidth]{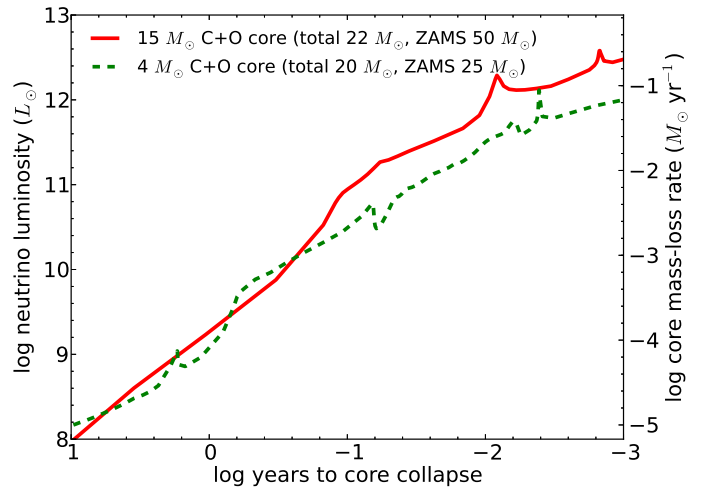}
\caption{
Neutrino luminosities of massive stars shortly before the core collapse.
The right-hand side vertical axis shows the corresponding core mass-loss
 rate which is estimated with Equation (\ref{neumasslossrate}).
The mass-loss rate from the stellar surface can be as high as the core
 mass-loss rate if the star is near the Eddington luminosity.
}
\label{neulumi}
\end{figure}

Figure \ref{neulumi} presents the neutrino luminosities obtained from the
stellar evolution models.
Most of neutrinos are thermally emitted and the luminosity increases
as the core temperature increases.
The obtained neutrino luminosity is consistent with those estimated from
other models \citep[e.g.,][]{odrzywolek2011,odrzywolek2004}.
We also show the corresponding mass-loss rate from the stellar core
due to the neutrino emission on the right-hand axis of Figure~\ref{neulumi}.
The total mass lost from the core by the neutrino emission
within a year before the core collapse is
$1.4\times 10^{-3}$ $M_\odot$ and $3.3\times 10^{-3}$ $M_\odot$
in the 25 \Msun\ and 50 \Msun\ models, respectively.

We present the stellar models to estimate the typical neutrino
luminosities of massive stars immediately before the explosion.
The Eddington factor $\Gamma$ of the 25 \Msun\ and 50 \Msun\ models
at the core collapse are 0.0033 and 0.98, respectively.
These stars need to be closer to the Eddington luminosity to enhance
mass loss by the central neutrino emission (see Section \ref{discussion}).
However, the central neutrino emission is expected to be
almost independent of the stellar
surface luminosity and the stars whose luminosities are high enough to
enhance mass loss should have similar neutrino luminosities to
those obtained in this section.

\begin{figure*}
\centering
\includegraphics[width=\columnwidth]{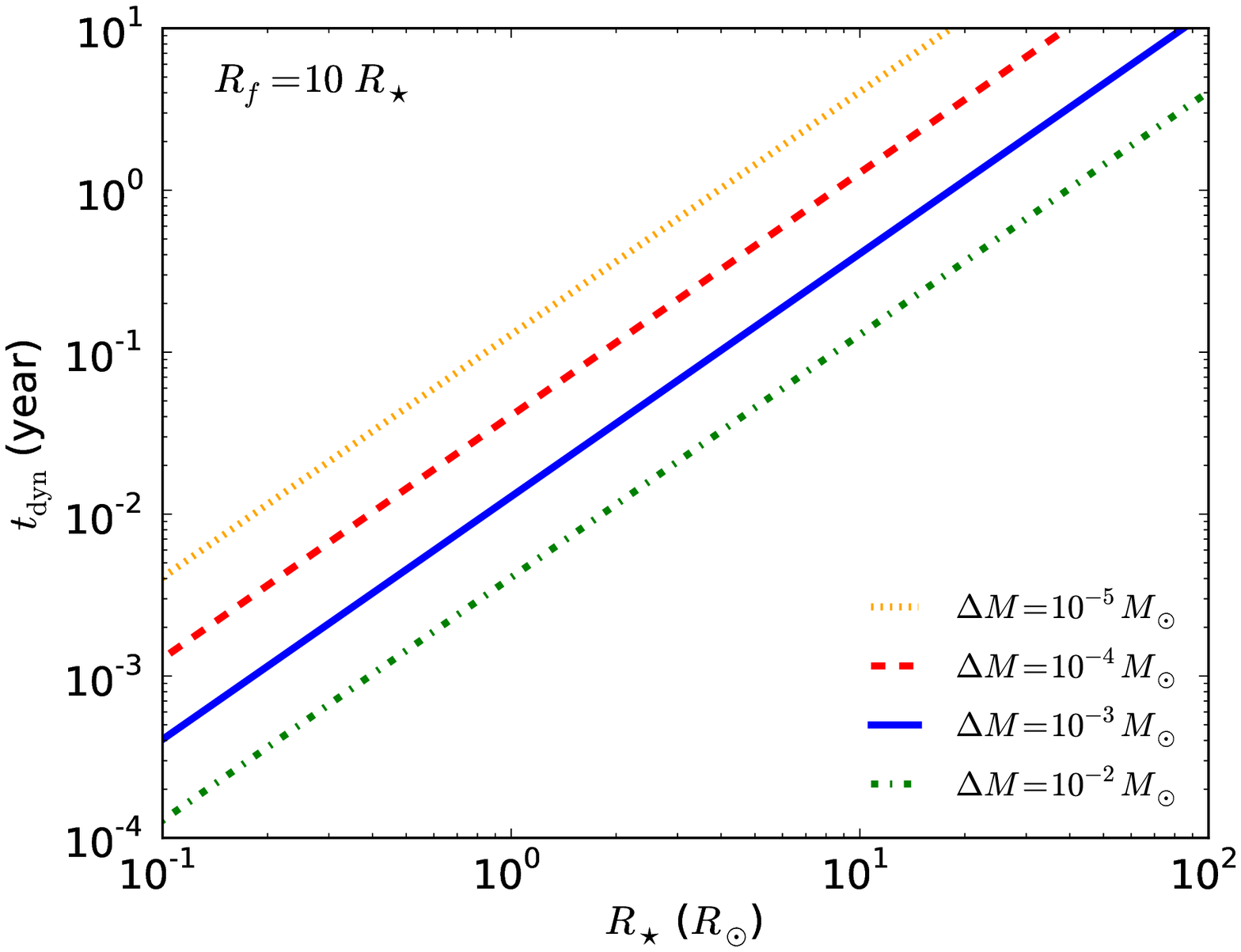}
\includegraphics[width=\columnwidth]{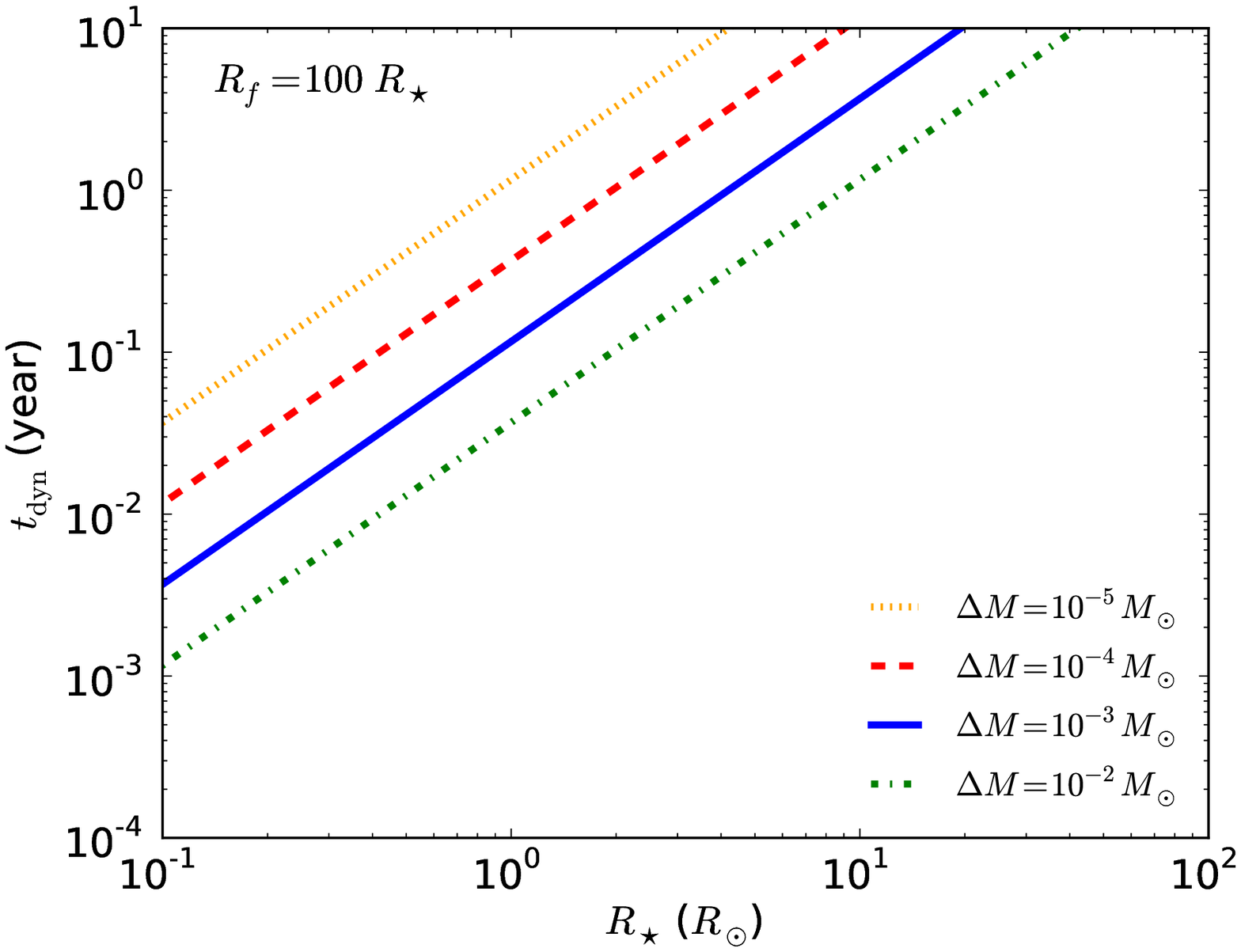}
\caption{
Dynamical time $t_\mathrm{dyn}$ of the material ejected from a stellar surface
because of the over-Eddington luminosity. $\Delta M$ is the mass lost
 from the stellar core. Left panel shows the dynamical time required
to reach $10~R_\star$ from $R_\star$. Right panel shows the dynamical
time required to reach $100~R_\star$ from $R_\star$.
}
\label{tdyn}
\end{figure*}

\section{Effect of neutrino mass loss on the stars near the Eddington luminosity}

Let us consider a star shortly before the core collapse
with mass $M_\star$, radius $R_\star$, and luminosity $L_\star$.
The Eddington luminosity $L_\mathrm{Edd}$ of the star is
expressed as
\begin{equation}
L_\mathrm{Edd}=\frac{4\pi GcM_\star}{\kappa},\label{Ledd}
\end{equation}
where $G$ is the gravitational constant and $\kappa$ is opacity.
Here, we consider the stars whose Eddington factor $\Gamma\equiv L_\star/L_\mathrm{Edd}$
is near unity $(\Gamma\approx 1)$.
If a stellar surface is at the Eddington luminosity, the gravitational force
and the radiation pressure are balanced. Thus, if the gravitational
force is suddenly reduced, the radiation pressure
gets larger than the gravitational force, initiating the stellar mass loss.
The Eddington luminosity is reduced linearly as
neutrinos escape (Equation \ref{Ledd}).

The resulting mass-loss rate from the stellar surface can be as high as
the central mass-loss rate by the neutrino emission as the 
gravitational energy released is available to push the surface.
This means that the star near the Eddington luminosity 
can have mass-loss rates as high as $10^{-4}$
$M_\odot~\mathrm{yr^{-1}}$ from about one year before
the explosion because of the central
neutrino emission (Figure \ref{neulumi}).
The line-driven wind mass-loss rates are typically below $10^{-4}$
$M_\odot~\mathrm{yr^{-1}}$ and the mass loss by the neutrino emission
can be dominant from about a year before the core collapse.
Overall, the mass-loss rate
increases as the star gets closer to the time of the core collapse.
The mass-loss rate can exceed $10^{-3}$ $M_\odot~\mathrm{yr^{-1}}$,
which is typical mass-loss rates of Type IIn SN progenitors,
from around $\sim$ 0.1 years before the core collapse.
The following central mass-loss rate
strongly depends on the core mass. 
It can exceed $10^{-2}$ $M_\odot~\mathrm{yr^{-1}}$ from $\sim$ 10 days
before the core collapse if the star has a massive core.

Since the high mass-loss rates by the super-Eddington luminosity
are achieved within about a year before the core collapse,
it is not trivial whether the surface material can actually escape far away
from the star within about a year.
Let us assume that $\Delta M$ is reduced from the core by the neutrino emission. 
As both the gravitational force and the radiative flux are 
inversely proportional to the square of the distance,
the motion of the material leaving the stellar surface is governed 
by the following equation of motion,
\begin{eqnarray}
\frac{d^2r}{dt^2}&=&-\frac{G\left(M_\star-\Delta M\right)}{r^2}+\frac{GM_\star\Gamma}{r^2},\\
&=&\frac{GM_\star}{r^2}\left(\frac{\Delta M}{M_\star}-1+\Gamma \right),\\
&\approx&\frac{G\Delta M}{r^2}.\label{eqmo}
\end{eqnarray}
$\Gamma\approx 1$ is assumed in deriving Equation (\ref{eqmo}).
For the moment, we neglect the time dependence of $\Delta M$ for
simplicity.
By integrating Equation (\ref{eqmo}), we can estimate the dynamical
timescale $t_\mathrm{dyn}(R_f)$ with which the material ejected from $R_\star$
reaches $R_f$,
\begin{equation}
t_\mathrm{dyn}(R_f)=\sqrt{\frac{R_\star^3}{2G\Delta M}}
\left(\frac{\sin\theta_f}{\cos^2\theta_f}+\ln\left[\tan\left(\frac{\theta_f}{2}+\frac{\pi}{4}\right)\right]\right),
\end{equation}
where
\begin{equation}
\theta_f = \arccos\left(\sqrt{\frac{R_\star}{R_f}}\ \right)\ \ \ \ (0\leq\theta_f<\frac{\pi}{2}).
\end{equation}
Figure \ref{tdyn} presents the dynamical timescales. The left-hand panel shows
$t_\mathrm{dyn}(10~R_\star)$ for given $R_\star$ and $\Delta M$ and
the right-hand panel shows $t_\mathrm{dyn}(100~R_\star)$.
We can see that material ejected from a star with $R_\star\sim R_\odot$, which is a
typical radius of WR stars, can easily reach above $10~R_\odot$ within
several days and above $100~R_\odot$ in $\sim$ 10 days.
Material ejected from larger stars require more time to travel.
Material ejected from blue supergiants or luminous blue variables (LBVs) ($R_\star \sim 10~R_\odot$)
is still able to reach beyond $10~R_\star$ within a year.
Material ejected from red supergiants with $R_\star\sim
100~R_\odot$ rather remains near the progenitor at the time of their explosion.

We have neglected the time dependence of $\Delta M$ in deriving
$t_\mathrm{dyn}$. In reality, $\Delta M$ increases with time.
Since the dynamical timescale becomes smaller as $\Delta M$
increases (Figure \ref{tdyn}), the previous arguments obtained by assuming
constant $\Delta M$ are presumed not to differ much even if
we take the time dependence into account.

Finally, the thermal adjustment timescale of the star,
\begin{eqnarray}
t_\mathrm{thm}&=&\frac{G M_\star^2}{R_\star \Gamma L_\mathrm{Edd}},\\
&=&
4800 \left(\frac{M_\star}{10~M_\odot}\right)
\left(\frac{R_\star}{R_\odot}\right)^{-1}
\left(\frac{\kappa}{0.2~\mathrm{cm^2~g^{-1}}}\right)
\Gamma^{-1}
~\mathrm{yrs},
\end{eqnarray}
is much larger than the dynamical timescales estimated above and the
material pushed by the exceeded radiative force can escape from the
system before the adjustment.

\section{Discussion}\label{discussion}
Some progenitors of Type IIn SNe are known to be near the Eddington
luminosity \citep[e.g.,][]{gal-yam2009,smith2011}. The large luminosity is often
related to LBVs which are known to experience
enhanced mass loss required to explain the observational properties of
Type IIn SNe \citep[e.g.,][]{smith2006}. However, LBVs are not SN progenitors in the
standard stellar evolution theory \citep[e.g.,][]{langer2012}.
Even if LBVs are actually SN progenitors,
the unknown mechanism which induces the extreme mass loss observed in LBVs
needs to be activated shortly before the core collapse by chance.
However, we suggest that, if a SN progenitor is near the Eddington
luminosity from several years before the core collapse, the star
is not required to be LBVs to enhance the mass loss.
The sudden luminosity increase observed in some Type IIn SNe
in $\sim$ years to
$\sim$ 10 days before the explosion is naturally expected from 
the high mass-loss rates of the stellar cores
caused by the large neutrino luminosity.

An important characteristic of the mass-loss mechanism presented here
is that it can make both hydrogen-rich and hydrogen-poor dense CSM.
It induces extreme mass loss shortly before the explosion of a star
independent of the surface composition.
The stellar luminosity need only to be close to the Eddington
luminosity.
There is growing evidence for the existence of hydrogen-poor dense CSM
around some SN progenitors.
Several Type Ib/c SNe show the signatures of the interaction with hydrogen-poor
dense CSM \citep[e.g.,][]{pastorello2007,gorbikov2013,ben-ami2013}.
Some Type Ic superluminous SNe are also suggested to be powered by the
interaction between dense hydrogen-poor dense CSM and SN ejecta
\citep[e.g.,][]{benetti2013,chatzopoulos2013,moriya2012,leloudas2012,quimby2011}.
However, the mechanisms to make the dense hydrogen-poor
CSM suggested so far
like the pulsational pair instability can only affect massive
stars in a limited range of mass
\citep[e.g.,][]{woosley2007,chatzopoulos2012}.
Our mechanism to enhance the mass loss can activate in a wider range
of mass. 
A WR star just need to be near the Eddington luminosity 
shortly before its explosion to have a very high mass-loss rate.

How close the star should be to the Eddington luminosity to activate
the mass loss due to the neutrino emission?
The Eddington factor $\Gamma$ needs to exceed 1
when the stellar core mass is decreasing. 
This means that $\Gamma$ should be larger than $1-\Delta M/M_\star$
when the neutrino luminosity becomes significant.
Assuming $\Delta M\sim 10^{-3}\ M_\odot$ and $M_\star\sim 10\ \Msun$,
$\Gamma\gtrsim 1-10^{-4}$ needs to be satisfied
shortly before the core collapse to enhance the mass loss by the neutrino emission.
Note that we are interested in massive stars shortly before the core collapse here.
If a massive star is born near the Eddington luminosity, it gets much
closer to the Eddington luminosity as it evolves
since it loses mass during its evolution by the normal stellar wind.
As the Eddington luminosity decreases with mass, the star can get much
closer to the Eddington luminosity at the
time of the core collapse. Thus, the required proximity to the Eddington
luminosity is plausible especially in the later stage of the star
we are interested in.

The mass loss caused by the central neutrino emission may prefer to be
activated in metal-rich environments.
This is because the opacity is expected to be higher in the metal-rich
environments. Then the Eddington luminosity can get lower and a star
may reach the Eddington luminosity more easily.
In addition, the normal wind mass loss which helps a star to reach the
Eddington luminosity is weakened in metal-poor environments.
However, the mass-loss mechanism by the core neutrino emission
itself can be activated regardless of
the stellar metallicity and it can enhance the mass loss
even in metal-poor environments.

If the star is $\Gamma< 1-\Delta M/M_\star$, material at the
stellar surface does not escape from the star. 
The reduced gravitational force by the neutrino emission may still affect
the stellar structure.
The star is presumed to expand because of the reduced binding force.
Thus, the stellar radii of SN progenitors can be larger than those
expected from the stellar evolution theory.
However, the reduced core mass is typically $\Delta M\sim 10^{-3}$
\Msun\ in our models and the amount of the surface expansion may be small.

\section{Conclusion}
We suggest a new mechanism to induce high mass-loss rates
shortly before the explosion of massive stars.
The neutrino luminosities of massive stars increase as they advance
the nuclear burning. Neutrinos escape freely from the stellar core
and take mass out of the core.
The core mass-loss rate by the neutrino emission becomes higher than
$10^{-4}$ $M_\odot~\mathrm{yr^{-1}}$ at about a year before the core collapse.
The core mass-loss rate becomes higher than $10^{-2}$ $M_\odot~\mathrm{yr^{-1}}$
in $\sim10$ days before the core collapse (Figure \ref{neulumi}).
If a star is close enough to the Eddington luminosity at that time,
the stellar luminosity can exceed the Eddington luminosity
by the core mass loss because of the reduced gravitational force.
Then, the radiative
force which was balanced by the gravitational force can push the
material at the stellar surface outward and drive strong mass loss.
As the reduced gravitational energy is available to push the surface meterial,
the mass-loss rate at the stellar surface can be as high as the core
mass-loss rate.
The extreme mass loss by the core neutrino emission can occur
even if the star does not have a hydrogen-rich envelope
as long as it is near the Eddington luminosity.
Thus, the existence of hydrogen-poor, as well as hydrogen-rich,
dense CSM recently observed in some SNe is naturally expected from this
mechanism. Even if a star is not close enough to the Eddington luminosity
to enhance the mass loss by the neutrino emission shortly before the
core collapse, it can still expand because of the reduced gravitational force.
This indicates that some stars 
may have larger radii at the time of the explosion
than those expected from the stellar evolution theory.

\begin{acknowledgements}
I would like to thank the referee for the comments which
improved this work significantly.
The author is supported by the Japan Society for the Promotion of Science
Research Fellowship for Young Scientists (23\textperiodcentered 5929).
Numerical computations were carried out on computers at Center for Computational Astrophysics, National Astronomical Observatory of Japan.
\end{acknowledgements}

\end{document}